\documentstyle[aps,prl,preprint]{revtex}

\begin{document}
\draft
\title{Conductance fluctuations near Anderson transition}
          
\author{Arindam Ghosh\footnote{email: aghosh@physics.iisc.
ernet.in}}
\address{Department of Physics, Indian Institute of Science, 
Bangalore 560 012, India}
\author{A. K. Raychaudhuri\footnote{email: arup@csnpl.ren.nic.in}}
\address{National Physical Laboratory, K.S. Krishnan Road, 
New Delhi 110 012, India}

\maketitle

\begin{abstract}
In this paper we report measurements of conductance fluctuations
in single crystal samples of Si doped with P and B close to the 
critical composition of the metal insulator transition 
($n_{c} \approx 4\times10^{18} cm^{-3}$). 
The measurements show that the noise, which arises from bulk sources, 
does not diverge as the Ioffe-Regal limit 
($k_{F}l \rightarrow$ 1) is approached from the metallic side. 
At room temperatures, the magnitude of the noise shows a shallow 
maximum around 
$k_{F}l \approx$ 1.5 and drops sharply as the insulating state    
is approached.
\end{abstract}
\vspace{1cm}

\newpage

Electron localization and Metal-Insulator (MI) transition has been a topic 
of considerable interest for quite some time and in particular in last two 
decades after the scaling theory clarified some of the key physics 
ingredients~\cite{G4}. 
One of the most researched mechanism of the MI transition is 
the Anderson-Mott transition that occurs in semiconductors doped to a      
critical concentration ($n_{c}$). 
A number of thermodynamic and transport  
studies have been  done in the past to understand the nature of the 
transition~\cite{LR,MILL}. 
One very important physical quantity that has not been 
investigated in doped semiconductors close to the critical composition is 
the conductance fluctuations or noise. In this paper, we report the results 
of the measurement of conductance noise in single crystals of Si doped with P 
and B so that we can approach the critical region from the metallic side 
($n/n_{c} \rightarrow$ 1).

In this paper we sought an answer to one important question: 
does the magnitude of conductance fluctuations 
diverge as we approach the Anderson transition in the 
heavily doped Si system? We think that this issue has not been 
rigorously looked into. The only reported  experiment that has 
systematically studied the fluctuations
as a function of disorder close to Anderson transition is in thin     
films of In$_{2}$O$_{x}$~\cite{COR}. 
The authors reported a sharp rise in the magnitude of the conductance 
noise (measured close to the room temperature) as the disorder is 
increased and the Ioffe-Regal limit ($k_Fl \approx$ 1) is approached.
It is thus of interest to investigate whether it is a universal 
phenomenon.

Conductance fluctuations with 
spectral power density $\propto 1/f$ (often known as $1/f$ noise) has been 
seen in disordered  
metallic films (like that in Ag and Bi)~\cite{DB} 
and also in oxide~\cite{AG} and C-Cu composites~\cite{MBW1} 
near the composition close to the M-I transition. The main 
focus of these works , based on disordered films, was to investigate 
the Universal Conductance Fluctuations (UCF)~\cite{FLS}. 
The issue of divergence (or its absence) of fluctuations  as a 
function of disorder has been investigated. However, no 
work  has been reported so far on experimental determination of 
conductance fluctuation in doped crystalline semiconductor (like Si doped   
with P or B) with concentration close to the       
critical composition. Our choice of 
doped single crystal Si was mainly guided by the fact that the 
Anderson transition has been most well studied in this system and most    
theoretical work has taken this as a model substance. This is also a well 
defined system in which it is possible to get well characterized 
samples.

Polished wafers  of  $\langle111\rangle$ 
orientation(grown by the Czochralski method) and  thickness $\approx 
300 \mu$m were sized down to a length of 2 mm, width 0.1-0.2 mm and were
thinned down   by etching to a thickness of 15-25 $\mu$m.   
(The samples used in this experiment were kindly supplied by 
Prof. D.H. Holcomb 
of Cornell University.) These wafers were used previously extensively in 
conductivity studies~\cite{UWE}. Details of growth and conductivity data 
can be found elsewhere. Table I contains the necessary numbers. 
In all, we investigated five 
different samples with $k_Fl$ varying between 2.8 to 0.78. Calculation of
$l$, the mean free path, is based 
on room temperature resistivity). This contains both uncompensated (Si(P)) 
and compensated (Si(P,B)) samples.

The noise was measured by a five probe ac technique~\cite{JS}  on  
samples of bridge type configuration with active volume for noise 
detection ($\Omega$) $\approx 10^{-6}$ cm$^3$
with peak current density $\sim 10^2$ 
A/cm$^2$. The noise was measured at $T$ = 300 K and $T$ = 4.2 K 
with temperature stability better than 10 mK. 
The background noise primarily
consisted of Johnson noise $4k_BTR$ from the sample. The spectral 
power density $S_{v}(f) \propto V^2_{bias}$. Leads of gold wires of 
diameter $\approx 25 \mu$m were bonded to the sample by a specially
fabricated wire bonder. The 
contacts were  Ohmic and have temperature independent contact 
resistance $\ll 1 \Omega$. All the relevant numbers about the      
samples studies are given in table I. In this table the 
mean free path $l$ in the parameter $k_Fl$ has been          
obtained from the room temperature resistivity data. The zero    
temperature conductivity, $\sigma_{0}$, shown in table I, has 
been obtained from the conductivity ($\sigma(T)$) below 4.2 K by 
using the power law, $\sigma(T) = \sigma_{0} + mT^\nu$.

For all the samples studied the spectral power density at a given frequency 
($\approx$ 1 Hz) was found to depend strongly with sample volume $\Omega$ 
when it was varied by more than a factor of 20. We show three examples in
figure 1. Typically, 
$S_v(f) \propto \Omega^{-\nu}$ with $\nu \approx 1.1-1.3$. 
This is seen at both 300 K and 4.2 K. This implies that 
predominant contribution to the noise arises from the bulk. A     
strong surface or contact contribution weakens the dependence of noise on 
$\Omega$ and makes $\nu < 1$. This is an important observation because in 
previous studies on semiconductors (done on films or devices with 
interfaces) the doping concentration was much smaller
($n \ll n_c$) and the 
noise had substantial contribution from surfaces or 
interfaces~\cite{MBW2}. Our 
experiments clearly show that the noise in heavily doped  single crystals
arises from the bulk.

In figure 2 we show the noise (measured at $f = 3$ Hz) as 
a function of the 
parameter $k_Fl$ for the 5 samples studied by us. The data 
at 4.2 K are shown in the inset.
Here $k_F$ is determined from the carrier 
density $n$ using $k_F = (3\pi^2n)^{1/3}$ and $l$ was 
determined from 
the room temperature resistivity $\rho$ using the free electron 
expression relation $l = \hbar k_F/ne^2\rho$.
The noise is expressed through the normalized value      
$\gamma$ defined as:

\begin{equation}
\gamma = fS_v(f)(\Omega n)/V^2_{bias}
\end{equation} 

\noindent In this 
representation we used $\gamma$ as a dimensionless number which 
represents the normalized noise. $\gamma$ is often referred to 
as the Hooge's parameter. Strictly speaking, this normalization to 
a frequency independent $\gamma$, is valid
only for $S_v(f) \propto 1/f$. To be consistent we have evaluated 
$\gamma$ at $f =$ 3 Hz for all the samples.
It can be seen in figure 2 that $\gamma$, has a distinct 
dependence on $k_Fl$. At $T = 300$ K $\gamma$ 
shows a shallow hump at   $k_Fl \approx$ 1-1.5. However as the   
insulating state is approached $\gamma$ shows a turn around and 
actually decreases. At $T = 4.2$ K 
(see inset of figure 2) $\gamma$ has a peak
at $k_{F}l \approx$ 2.3. However, $\gamma$ stays close to 1
and does not diverge as $k_Fl \rightarrow$ 1.
The mechanism of noise at 4.2 K and 300 K are expected to be different.
As a result we do not 
expect the same dependence of $\gamma$ at two widely different 
temperatures. However, our data shows that irrespective of the temperature,
$\gamma$ does not diverge as we enter the insulating state.
This is unlike what has been seen in disordered films of  
In$_{2}$O$_{x}$  where $\gamma > 10^5$ when $k_Fl \leq$ 1.
For the sake of comparison this is shown in figure 
3 along with our data. In our case $\gamma$ never becomes as large as 
10$^{5}$-10$^{7}$ as seen in the oxide films and  over the whole 
range $\gamma$ is substantially smaller. In the same graph we have
shown $\gamma$ of thin metal films.
For Si(P,B) samples the $\gamma$
are at least three orders of magnitude higher 
than that seen in conventional thin metallic films 
($\gamma \approx 10^{-3}$ to 10$^{-5}$). It is extremely interesting to   
note that lightly doped Si films on sapphire samples 
$\gamma \approx 10^{-3}$ although in this case it is likely that 
the noise arises from the surfaces/interfaces. The lightly doped 
Si samples in which $\gamma \approx 10^{-3}$ has been observed, 
the doping level $\approx 10^{13}-10^{14}$ /cm$^{3}$. In our 
case the sample which has the least $\gamma$ at room temperature 
has a level of 
doping $\approx 4\times10^{18}$ /cm$^{3}$ and for this sample $\gamma$ 
is already down to 0.25. If this trend continues then 
$\gamma \rightarrow 10^{-3}$ for $n \leq 10^{16}$ /cm$^{3}$. We 
believe that for $n$ less than this level of doping, the surface 
states will dominate the noise mechanism.

We next investigate the spectral dependence of the noise power 
$S_v(f)$. At both $T =$ 4.2 K and 300 K the predominant spectral 
dependence is almost $1/f$ type with $S_v \propto 1/f^\alpha$ 
with $\alpha \approx$ 0.9-1.25. This $1/f$ dependence has been 
seen over six orders of magnitude in 
frequency in the range $f \sim 10^{-4}$ Hz to 10$^2$ Hz for 
three samples with $k_Fl =$ 2.8 (PS24), 1.68 (D150) and 0.78 (E90). At 
$T =$ 4.2 K the spectral dependence of noise tend to deviate from
pure $1/f$ form. This can be seen 
in figure 4, where we have plotted the data as $f\times S_{v}(f)$ vs. 
$f$. For all the samples the $f\times S_v(f)$ is featureless at room 
temperature and the slope corresponds to 
$\alpha \approx$ 1.05-1.2. At $T =$ 4.2 K, the uncompensated (and more 
metallic) samples retain their $1/f$ form. However, as the 
disorder increases on compensation by B doping, additional 
features show up as can be seen in figure 4. This is quite 
prominent in the most disordered sample E90 which is rather close to 
the insulating side ($\sigma_0$ = 0). Given the scope of the     
paper we do not elaborate on this point. However, we note the     
important observation that as the critical region is approached 
($k_Fl \rightarrow$ 1), the spectral dependence of the noise 
power undergoes a change.

As pointed out earlier, the dependence of $\gamma$ on $k_Fl$ 
at both $T =$ 4.2 K and 300 K is drastically different from that 
carried out on thin disordered films of In$_{2}$O$_{x}$ near the 
Anderson transition~\cite{COR}. Study of conductance fluctuations
noise near Anderson transition has also been done in 
C-Cu films~\cite{MBW1}. Interestingly, in this case the $\gamma$ 
$\approx$ 1-5 at room temperature 
for all the samples which are close to the 
critical region. The noise at $T =$ 4.2 K in the same systems show 
some what larger $\gamma \approx$ 10 but it is not as large as 
that seen in the In$_{2}$O$_{x}$ films. Also, noise measurements near
the Anderson transition in  La$_{1-x}$Sr$_x$VO$_3$ thin films did not show
any indication of divergence~\cite{SM}.
Another study of noise where the            
metal-insulator boundary has been crossed is the investigation    
on a percolating system Pt/SiO$_{2}$ composite done at 300 K~\cite{REF}. In  
this case, however, $\gamma$ undergoes a change by 3-4 orders of 
magnitude when the percolation threshold is crossed.
We can then conclude that the behavior of $\gamma$ close to the 
critical region of Anderson transition may not be universal. In all 
likelihood, it depends on the mechanism that produces the 
noise. For the Si(P.B) we are carrying out an extensive 
investigation of the noise to find the temperature and field 
dependence which can identify the mechanism which is causing the 
noise. This will be elaborated in a future publication.

\newpage
\begin{table}[t]\caption{}
\begin{center}
\vspace{0.1cm}
\begin{tabular}{|l|c|c|c|c|c|}\hline
\label{tab1}
Sample (dopant)& $n$ (cm$^{-3}$)      & K   & $n/n_c$ & $\sigma_{RT}$ (S/m) & $k_Fl$ \\ \hline\hline
PS24 (P)       & 1.0$\times10^{19}$   & -   & 2.5     & 1.5$\times$10$^4$   & 2.80\\ \hline
PS41 (P)       & 6.5$\times10^{18}$   & -   & 1.5     & 1.1$\times$10$^4$   & 2.30\\ \hline
D150 (P,B)     & 1.0$\times10^{19}$   & 0.4 & 2.0     & 0.9$\times$10$^4$   & 1.68\\ \hline
C286 (P,B)     & 5.5$\times10^{18}$   & 0.5 & 1.1     & 3.7$\times$10$^3$   & 0.84\\ \hline
E90  (P,B)     & 4.5$\times10^{18}$   & 0.6 & 1.0     & 3.3$\times$10$^3$   & 0.78\\ \hline
\end{tabular}
\end{center}
\end{table}

\newpage

{\bf\Large Figure Caption}

\vspace{0.5cm}
{\bf Fig.1.} Volume ($\Omega$) dependence of the noise in 
Si(P,B) samples at room temperature. Similar dependence 
has been observed at $T =$ 4.2 K as well. Relatively large
error in volume determination results from rounding-off of 
the edges of the samples during chemical etching. The dotted line
has a slope of $\approx$ 1.1.
 
\vspace{0.5cm}
{\bf Fig.2.} Variation of the normalized noise parameter 
$\gamma$ as a function of disorder as measured by the parameter 
$k_Fl$ at $T =$ 300 K. Inset shows that data at $T =$ 4.2 K.
The solid line is guide to the eye.

\vspace{0.5cm}
{\bf Fig.3.}  Comparison of $\gamma$ for different solids with 
varying $k_Fl$. The data points show our data. The shaded and 
hatched regions have been taken form other published data, 
principally ref.5. 

\vspace{0.5cm}
{\bf Fig.4.}  Variation of spectral density of noise with        
frequency at $T =$ 300 K and $T =$ 4.2 K in three 
representative samples. Data for different samples have been
shifted for clarity.

\begin{thebibliography}{1-30}
\bibitem{G4} Abrahams E, Anderson P W, Licciardellow D C and Ramakrishnan 
T V 1979 Phys Rev Lett {\bf 42} 673
\bibitem{LR} Lee P A and Ramakrishnan T V 1985 Rev Mod Phys {\bf 57} 287 
\bibitem{MILL} Milligan R F, Rosenbaum T F, Bhatt R N and Thomas G A 1985
in {\it Electron-Electron Interaction in Disordered Systems} ed. Efros A and
Pollak M p.231 and references therein
\bibitem{FLS} Lee P A, Stone A D and Fukuyama H 1987 Phys Rev B {\bf 35} 1039
\bibitem{COR} Cohen O, Ovadyahu Z and Rokni M 1992 Phys Rev Lett {\bf 69} 
3555 
\bibitem{DB} Dutta P and Horn P 1981 Rev Mod Phys {\bf 53} 497; Birge N O,
Golding B and Haemmerle 1990 Phys Rev B {\bf 42} 2735
\bibitem{AG} Cohen O and Ovadyahu 1994 Int J Mod Phys B {\bf 8} 897 
\bibitem{MBW1} Garfunkel G A, Alers G B, Weissman M B, Mochel J M and 
VanHarlingen D J 1988 Phys Rev Lett {\bf 60} 2773
\bibitem{UWE} Thomanschefsky U H 1990 PhD thesis, Cornell University 
\bibitem{JS} J.H. Scofield 1987 Rev Sci Instrum {\bf 58} 985 
\bibitem{MBW2} Weissman M B 1988 Rev Mod Phys {\bf 60} 537; Hooge N, 
Kleinpenning T.G.M. and Vandamme L.K.J. 1981 Rep. Prog. Phys. {\bf 44} 479 
\bibitem{SM} Prasad E, Sayer M and Noad J P 1979 Phys Rev B {\bf 19} 5144
\bibitem{REF} Mantese J V, Curtin W A and Webb W W 1986 Phys Rev B {\bf 33}
7897
\end{thebibliography}
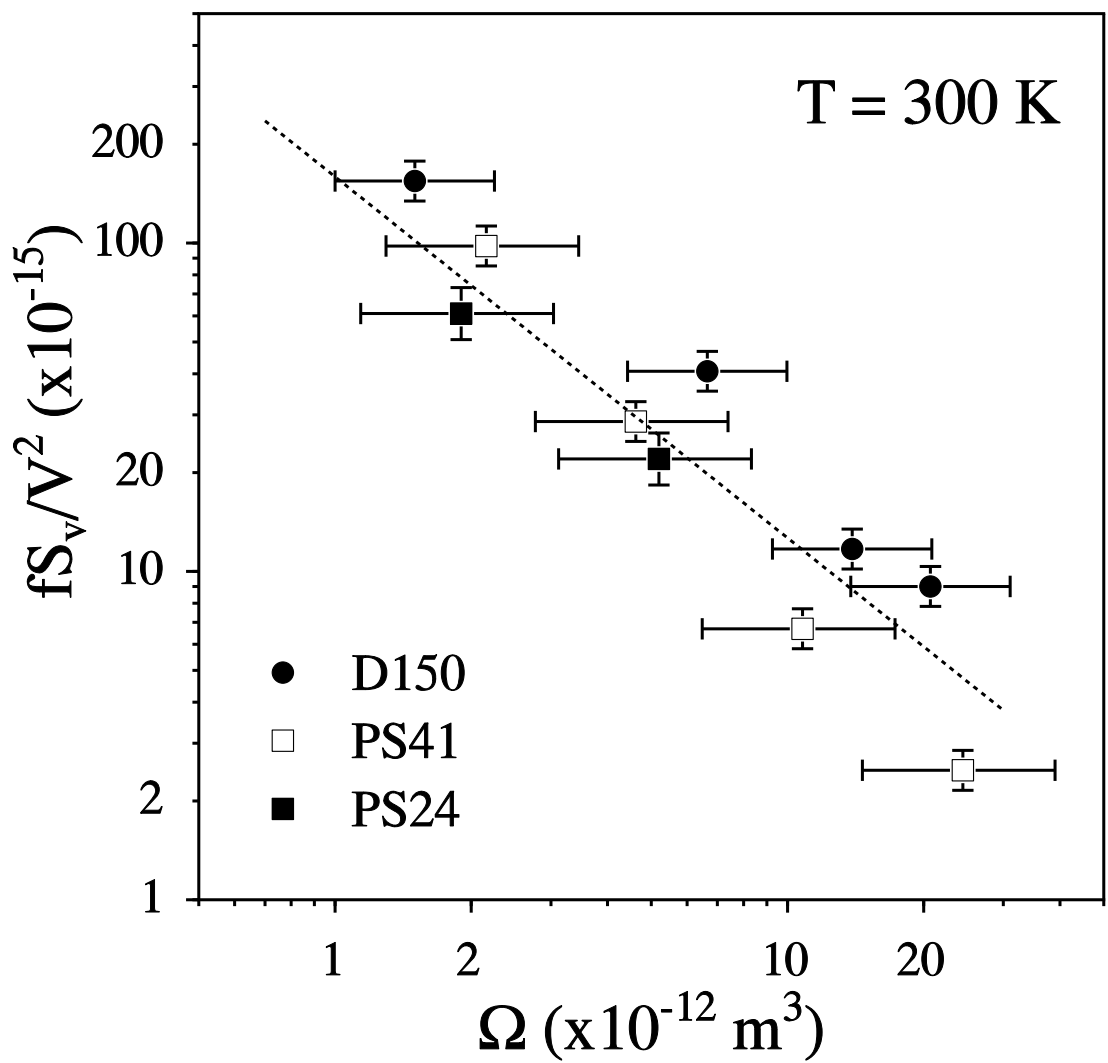
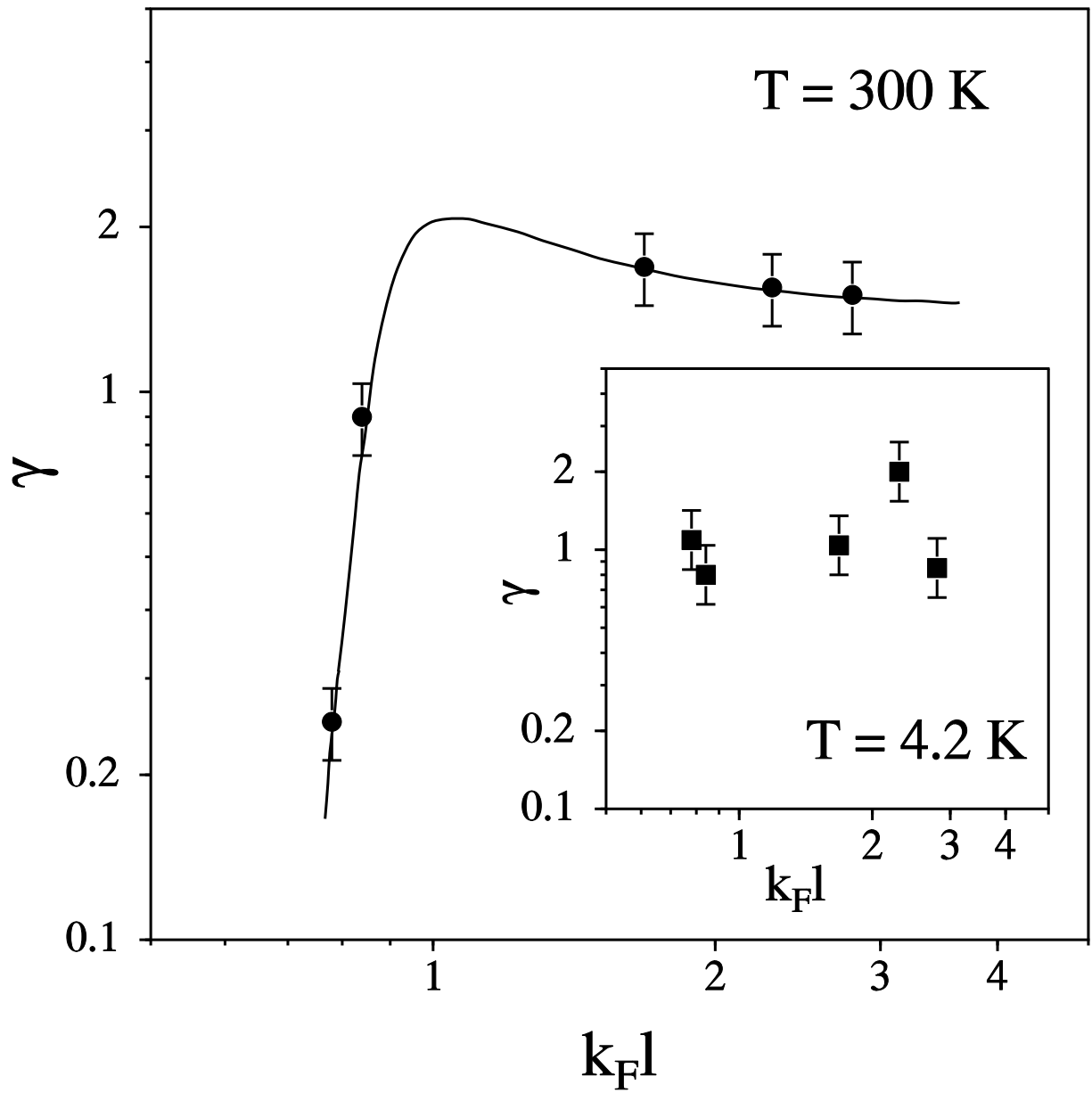
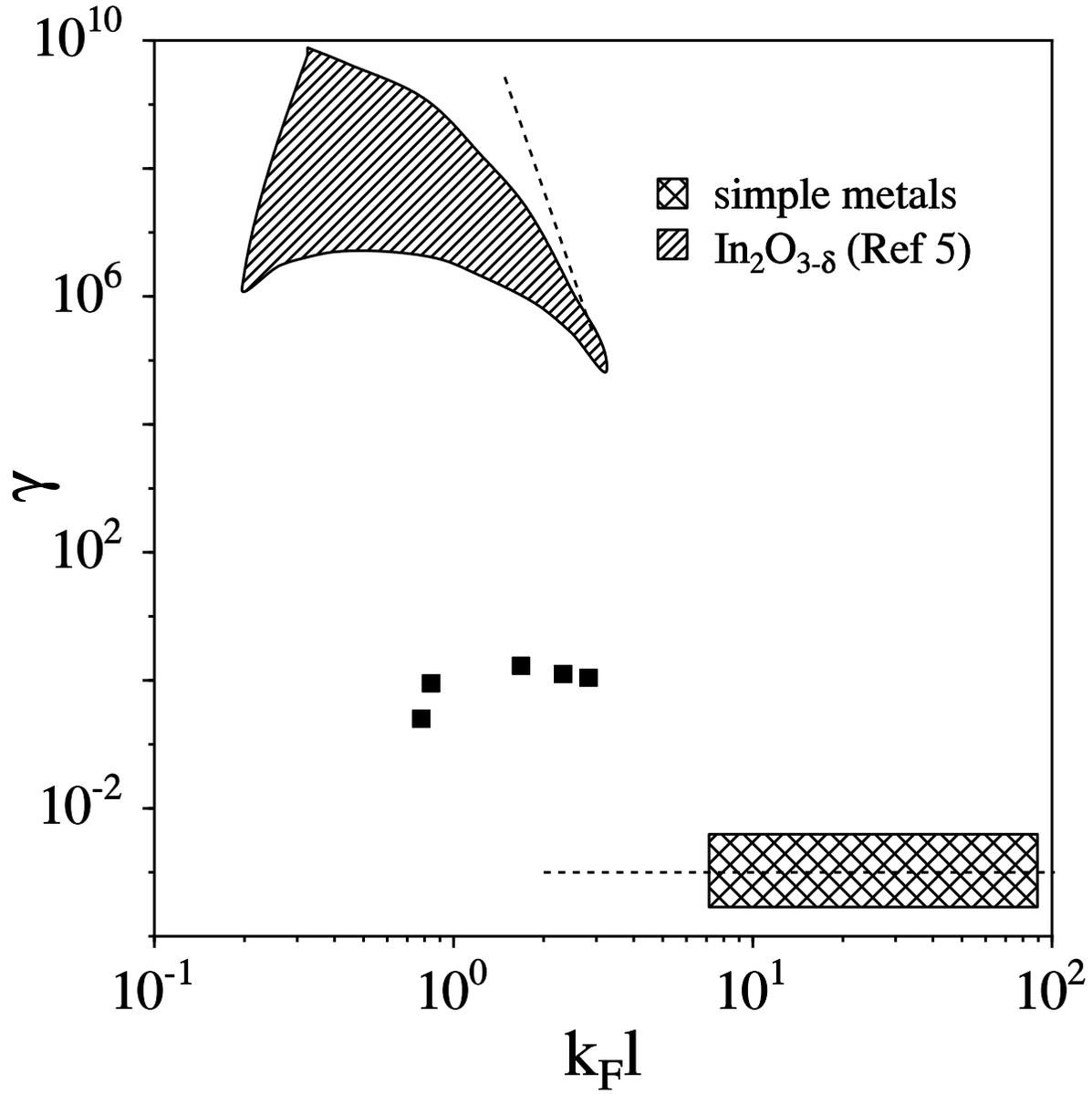
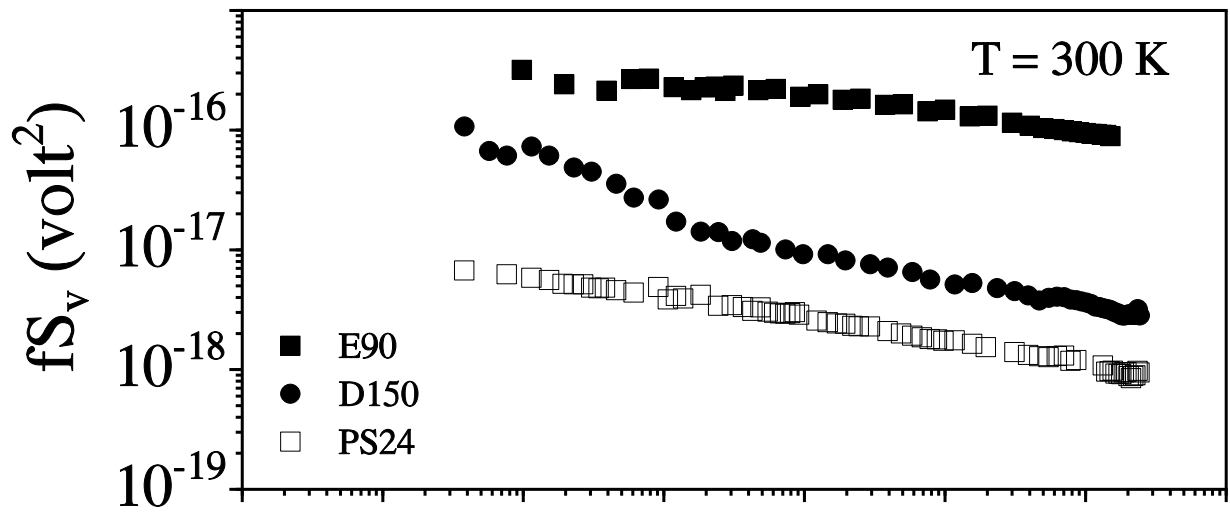
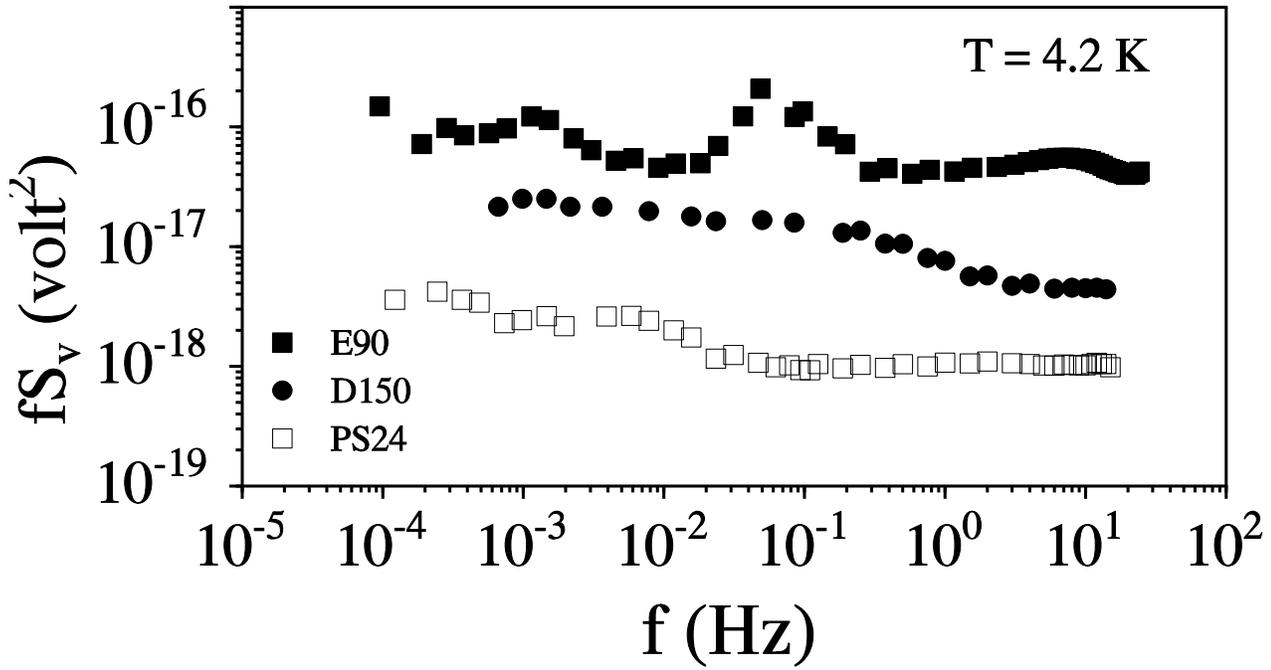
\end{document}